\documentstyle[epsfig,twoside,12pt]{article}
{\catcode `\@=11 \global\let\AddToReset=\@addtoreset}
\AddToReset{equation}{section}

\def\greaterthansquiggle{\raise.3ex\hbox{$>$\kern-.75em\lower1ex\hbox{$\sim$}}}
\def\lessthansquiggle{\raise.3ex\hbox{$<$\kern-.75em\lower1ex\hbox{$\sim$}}}

\newcommand{\beq}{\begin{equation}}
\newcommand{\eeq}{\end{equation}}
\newcommand{\beqa}{\begin{eqnarray}}
\newcommand{\eeqa}{\end{eqnarray}}
\newcommand{\beqan}{\begin{eqnarray*}}
\newcommand{\eeqan}{\end{eqnarray*}}
\newcommand{\ba}{\begin{array}}
\newcommand{\ea}{\end{array}}
\newcommand{\no}{\nonumber}
\newcommand{\vect}{\overrightarrow}

\newcommand{\dar}{\downarrow}
\newcommand{\uar}{\uparrow}

\newcommand{\cP}{{\cal P}}
\newcommand{\R}{{\cal R}}

\newcommand{\Tr}{\rm Tr\,}

\newcommand{\Un}{\underline}

\newcommand{\ra}{\rightarrow}

\newcommand{\dfrac}{\displaystyle \frac}

\normalsize
\begin{document}
\bibliographystyle{plain}
\begin{titlepage}
\begin{flushright}
CERN-PH-TH/2006-260\\
ESI-1880(2006)\\
December 12, 2006
\end{flushright}

\vspace*{1.6cm}
\begin{center}
{\Large \bf High-$T_c$ superconductivity by phase cloning}
\\[40pt]

Nevena Ilieva$^{1,2 \, \ast}$ and Walter Thirring$^{2,3}$\\ [36pt]
{\small
\begin{tabular}{cl}
\qquad & $^1$ Theory Division, CERN, CH-1211 Geneva 23,
Switzerland\\[4pt]
& $^2$ Erwin Schr\"odinger International
Institute for Mathematical Physics\\
&  Boltzmanngasse 9, A-1090 Vienna, Austria\\[4pt]
& $^3$ University of Vienna, Institute for Theoretical
Physics\\
& Boltzmanngasse 5, A-1090 Vienna, Austria\\[12pt]
& E-mail: nilieval@mail.cern.ch
\end{tabular}}

\vspace{2.7cm}
\begin{abstract}

We consider a BCS-type model in the spin formalism and argue that
the structure of the interaction provides a mechanism for control
over directions of the spin $\vect S$ other than $S_z$, which is
being controlled via the conventional chemical potential. We also
find the conditions for the appearance of a high-$T_c$
superconducting phase.


\medskip

\end{abstract}
\end{center}

\vspace{1.8cm}

{\footnotesize

$^\ast$ On leave from Institute for Nuclear Research and Nuclear
Energy, Bulgarian Academy of Sciences, Boul.Tzarigradsko Chaussee
72, 1784 Sofia, Bulgaria}

\vfill
\end{titlepage}

\setcounter{page}{2}

\section*{Introduction}

Twenty years after the discovery of high-temperature superconductivity
\cite{BM} there is still neither consensus nor clear understanding of the
mechanism or mechanisms which are behind this exciting and with innumerable
practical applications phenomenon.The initial discussions (see, e.g.
\cite{Enz})
have led to the formation of some main conceptual stream, as presented in
\cite{A},
however other view points are continuously being argued, just to mention
a recent
one \cite{b-up}.

In \cite{NPB} we proposed a combination of a BCS and a mean-field
Hamiltonian where the transition temperature could become
arbitrarily high. This happened without increasing the interaction
indefinitely but by a small denominator. In this note we
investigate this effect more closely and find that another
important ingredient is a chemical potential which breaks the
electron conservation. We give a model for this phenomenon by the
interaction with a reservoir of quasi-particles which do not have a
definite electron number. Since such objects play an important
role in the theory of the Josephson currents \cite{Jos}, we think that this
possibility is not purely academic.

To avoid a terminological misunderstanding, we recall that
the (quantum-mechanical) mean-field theory and the BCS-theory
of superconductivity correspond to
essentially different physical situations. A
mean-field theory means that the particle density $\rho(x) =
\psi^*(x)\psi(x)$ (in second quantization) tends to a {\it c}-number
in a suitable scaling limit. With an appropriate
smearing, from the operator-valued distribution
$\rho(x)$ an unbounded operator is being produced, so that the best
to be strived
for remains the strong resolvent convergence in a representation
where the macroscopic density is built in. In the BCS-theory pairs
of creation operators with opposite momentum
$\tilde\psi^*(k)\,\tilde\psi^*(-k)$ tend to {\it c}-numbers, so the
correlations required in both cases, seem to be quite different.
The main result in \cite{NPB} was that both types of
correlations may well co-exist in certain regions of the parameter
space  (temperature, chemical potential, relative values of the two
coupling constants) and this appears to be the case in the
KMS-state of the equivalent approximating (Bogoliubov)
Hamiltonian  $H_B$, two Hamiltonians being considered as equivalent
if they lead to one and
the same time evolution of the local observables \cite{TW}.

In what follows, we generalize the original BCS model in the most natural
way, namely by augmenting it with the missing mean-field interaction
components. We show that this provides a mechanism for control
over directions of the spin $\vect S$ other than $S_z$, which is
being controlled via the conventional chemical potential.

\section{The degenerate BCS Hamiltonian}

The initial quartic BCS Hamiltonian is mainly known in terms of
fermionic creation and annihilation operators \cite{BCS} \beq H = \sum_k
(\omega_k-\mu)(a_{\uar,k}^\dag a_{\uar,k} + a_{\dar,k}^\dag
a_{\dar,k}) +
\sum_{k,k'}V_{k,k'}a^\dag_{\dar,k'}a^\dag_{\uar,-k'}a_{\uar,-k}a_{\dar,k}.\eeq
It involves however only the algebra generated by the
pair-operators $a_{\uar,-k}a_{\dar,k}$ (observe
$a^\dag_{\uar}a_{\uar} + a^\dag_{\dar}a_{\dar} =
[a^\dag_{\uar}a^\dag_{\dar}, a_{\dar}a_{\uar}] + 1$), so we shall
only be concerned with them and shall represent them by spin
matrices $a_{\uar,j}^\dag a_{\dar,-j}^\dag\rightarrow\sigma_{j+} =
(\sigma_{j,x} + i \sigma_{j,y})/2$, $j=1, \dots, N$.
As a weak interaction can only scratch the Fermi surface, we take
$\omega_k=\omega$, $\forall k$ and incorporate the latter into $\mu$.
Finally,
we set $V_{k,k'} = -2\lambda_B/N \,\,\, \forall k, k'$. With the
notation $\vect S = \sum_{i=1}^N \vect {\sigma_i}$, Hamiltonian
(1.1) becomes equivalent to \beq H = -\frac{\lambda_B}{2N}(S_x^2 +
S_y^2) - \mu S_z.\eeq

In this form the Hamiltonian can be exactly diagonalized and the
following steps are mathematically rigorous in the limit.

We assume that the thermal state $\langle A \rangle = \Tr
Ae^{-\beta H}/\Tr e^{-\beta H}$ is such that the length of
$\vect S$ is much bigger than the fluctuations
around it, $\langle (\vect S - \langle \vect S\rangle )^2\rangle$.
This means that in the identity
$$
S^2 = (S-\langle S\rangle)^2 + 2\langle S\rangle S - \langle
S\rangle^2
$$
the first term is small compared to the second one. Since the last
term is a $c$-number, Hamiltonian (1.2) becomes equivalent to the
following one, linear in $\vect S$
\beq H_B =
-\lambda_B\left(\frac{\langle S_x \rangle}{N}
S_x+\frac{\langle S_y \rangle}{N} S_y\right) - \mu
S_z.\eeq Of course, the original Hamiltonian is invariant under
rotations around the $z$-axis, but the spin-vector $\vect S$ will
point into some direction (with $-\mu S_z$ contribution taken into account)
and we shall call
this resulting spin direction $S_B$, that is Eq. (1.3) can be rewritten as
\beq
H_B = -\lambda_B \frac{\langle S_x \rangle}{N} S_x -\mu S_z =:
-W_0 S_B,\eeq
where
\beqa
& W_0 = \sqrt{(\lambda_B\langle
S_x\rangle/N)^2
+ \mu^2}> |\mu| \\[8pt]
& S_B = bS_x + \sqrt{1-b^2}S_z, 
\eeqa
and the mixing parameter of the Bogoliubov transformation \cite{Bog} is
defined through
\beq b = \lambda_B\langle S_x\rangle /
NW_0.\eeq

This rotation can be inverted and if $S_\perp$ is in the $x-z$
plane orthogonal to $S_B$, we have (Figure 1)
$$
S_x = bS_B - \sqrt{1-b^2}S_\perp.
$$

\begin{figure}[htb]
\setlength{\unitlength}{1pt}
\begin{picture}(175,170)
\put(140,-10){\makebox(0,0)[lb] 
{\includegraphics*[145,40][320,195]{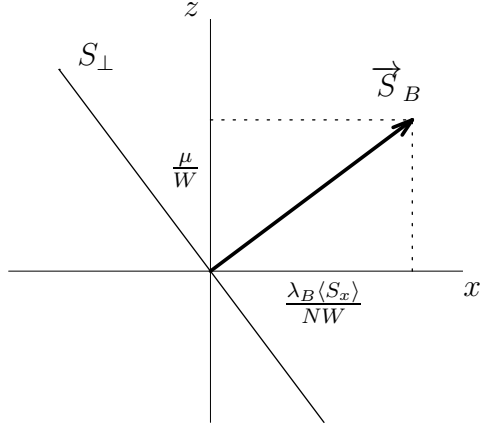}}}
\put(313,42){\makebox(0,0)[l]{$x$}}
\put(207,148){\makebox(0,0)[l]{$z$}}
\put(278,120){\makebox(0,0)[l]{$\vect S_B$}}
\put(167,130){\makebox(0,0)[l]{$S_\perp$}}
\put(202,87){\makebox(0,0)[l]{$\frac{\mu}{W}$}}
\put(245,37){\makebox(0,0)[l]{$\frac{\lambda_B\langle S_x\rangle}{NW}$}}
\end{picture}
\begin{quote}
\caption{The total-spin decomposition.}
\end{quote}
\end{figure}

Since $H=-W_0S_B$ is the sum of $N$ spins in the $B$-direction, Eq.(1.4),
the thermal expectation values are the usual ones
\beq \frac{\langle S_B \rangle}{N} = \tanh
\frac{W_0}{2T},\qquad \langle S_\perp \rangle = 0. \eeq The
self-consistency of the system is expressed by the so-called
``gap-equation" \beq \frac{\langle S_x\rangle}{N} =
b\tanh \frac{W_0}{2T} = \frac{\lambda_B}{W}
\frac{\langle S_x \rangle}{N} \tanh \frac{W_0}{2T}.\eeq

This gap-equation has two solutions
\begin{itemize}
\item[{\bf (A)}] {\sl a normal state} \beq\langle S_x\rangle = 0
\quad \forall T \eeq \item[{\bf (B)}] {\sl a superconducting
state} \beqa\langle S_x\rangle \not= 0, \quad \frac{W_0}{\lambda_B}=
\tanh\frac{W_0}{2T},\\[6pt]
\mbox{for } T =
\frac{\lambda_B}{2}F\left(\frac{W_0}{2T}\right)<T_c,\eeqa
\end{itemize}

\begin{figure}[htb]
\setlength{\unitlength}{1pt}
\begin{picture}(235,150)
\put(100,-10){\makebox(0,0)[lb] 
{\includegraphics*[90,0][325,150]{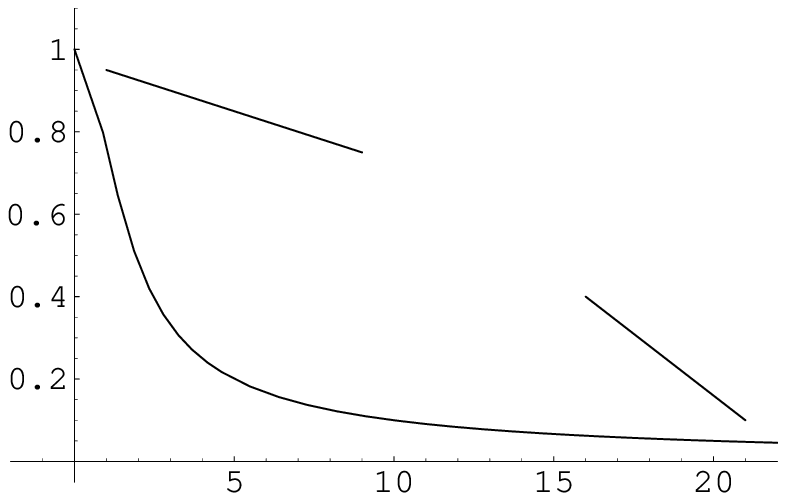}}}
\put(85,145){\makebox(0,0)[l]{$F(\alpha)$}}
\put(327,-6){\makebox(0,0)[l]{$\alpha$}}
\put(100,-6){\makebox(0,0)[0]{$1$}}
\put(175,85){\makebox(0,0)[l]{$\mbox{{\sl high-T region}}$}}
\put(238,62){\makebox(0,0)[l]{$\mbox{{\sl low-T region}}$}}
\end{picture}
\begin{quote}
\caption{The characteristic function $F(\alpha)$: high-T region
corresponds to small $\alpha$.}
\end{quote}
\end{figure}

\noindent where the characteristic function $0<F(\alpha)\leq 1$ is given by
$$
F(\alpha) = \frac{\tanh \alpha}{\alpha}\,\ba{ccll} & &
1-\alpha^2/3, & \alpha\ra 0\\
\nearrow & & & \\
\searrow & & & \\
& & 1/\alpha, & \alpha\ra\infty \ea
$$
(see Figure 2) and the critical value which the temperature $T$ for no values of $\mu$ and $\lambda$ can exceed is
\beq
T_c
\leq\frac{|\lambda_B|}{2}.\eeq

On Figure 3, the pure
BSC-situation is shown: the plot of both sides of Eq.(1.11), for $T=\lambda_B/4,
\lambda_B/2, 3\lambda_B/4$. The limit (1.13) becomes obvious.

\begin{figure}[htb]
\setlength{\unitlength}{1pt}
\begin{picture}(235,150)
\put(100,-10){\makebox(0,0)[lb] 
{\includegraphics*[90,0][325,150]{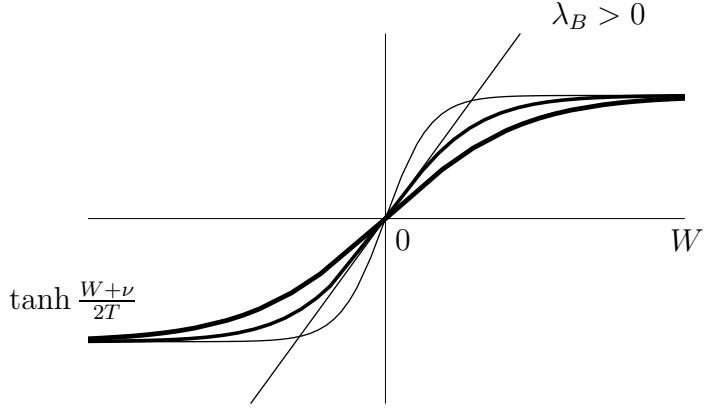}}}
\put(70,32){\makebox(0,0)[l]{$\tanh\frac{W+\nu}{2T}$}}
\put(320,55){\makebox(0,0)[l]{$W$}}
\put(216,55){\makebox(0,0)[l]{$0$}}
\put(275,140){\makebox(0,0)[l]{$\lambda_B>0$}}
\end{picture}
\begin{quote}
\caption{The pure BCS situation: plot of both sides of Eq.(1.11), for
$T=\lambda_B/4, \lambda_B/2, 3\lambda_B/4$ (the line thickness increases with $\lambda_B$).}
\end{quote}
\end{figure}

Relations (1.5), (1.12) in fact suggest some possibilities for high-$T_c$
generation. A realistic mechanism should result in a deviation from the
proportionality relation (1.12) as far as the characteristic-function part is
considered (as $F(\alpha)$ is always less than 1). Also,
it should aim a modification of the quasiparticle dispersion relation (1.5),
as is e.g. the case of the gossamer superconductor \cite{Laugh}.
We rather target
the appearance of an effective chemical potential, whose
variation would provide a means of influence on the transition temperature.

\bigskip

\section{On the role of the chemical potential}

The chemical potential is a control parameter which adjusts the
number of Cooper pairs, in our formalism $S_z$. We shall now argue
that the BCS-interaction gives us a handle to control also other
directions of $\vect S$ and in this way to clone the Josephson phase of $R$.

Suppose our system interacts strongly
with a superconducting reservoir $R$ such that Eq.(1.3) holds for the
ensemble:
\beq
H=-\frac{\lambda_B}{2}\left((\vect S + \vect S^{(R)})
\frac{\langle\vect S + \vect S^{(R)}\rangle}{N + N_R} - (S_z +
{S_z}^{(R)})\frac{\langle S_z + {S_z}^{(R)}\rangle}{N +
N_R}\right).\eeq

\noindent The cross-term \beq \frac{\vect S\langle \vect
S^{(R)}\rangle}{N+N_R} \eeq induces 
control
parameters which just copy on the system the situation in $R$, if
$R$ is dominant \cite{TW, T, HN}.
By the coupling with the reservoir, the $\vect S$-direction is cloned. We shall
call this spin-coaxial exchange. Thus,
if the reservoir is in the normal state, $\vect S^{(R)}$
is in the $z$-direction and we get the usual
chemical potential. If $R$ is superconducting, we get a
chemical-potential coupled $S_B^{(R)}$ (a $\nu S_B^{(R)}$-term in the
Hamiltonian), which represents the
quasi-particles --- the elementary excitations of the
superconductor. If the reservoir dictates a Josephson phase, e.g. in the $x$-direction, this
means that in the cross-term (2.2) $\langle S_x^{(R)}\rangle \sim \nu$.
Correspondingly, if we treat the Hamiltonian with this additional term as before, it becomes
\beq H-\nu S_x = -WS_B,\quad \nu>0\eeq with dispersion relation
$$W = \sqrt{\left(\nu + \lambda_B\dfrac{\langle
S_x\rangle}{N}\right)^2 + \mu^2}$$ and mixing parameter $$ c =
\dfrac{\nu + \lambda_B\langle S_x\rangle/N}{W}.$$

The gap-equation reads: \beq \frac{\langle S_x\rangle}{N} =
\frac{\nu + \lambda_B\langle S_x\rangle/N}{W}\,\tanh
\frac{W}{2T}\eeq and the temperature at which the gap opens becomes \beq T =
\frac{1}{2}\left(\frac{\nu}{\langle S_x\rangle/N} +
\lambda_B\right) F(\frac{W}{2T}).\eeq

So with this choice

{\bf (i)} expectedly, there is no normal phase, i.e. solution with
$\langle S_x\rangle = 0$;

{\bf (ii)} $\forall T$ there exists a solution with $\langle
S_x\rangle >0$.

Thus there is no phase transition $\forall \, \nu >0$. It is quenched, since
the symmetry is broken externally and not spontaneously.

\medskip

With all this taken into account, for the description of a system which
exhibits a phase transition towards a high-temperature superconducting phase, we study the following system: It has a BCS interaction of the form of (1.2) and a self-interaction $\sim \lambda_M S_z^2$ which can also be treated as a mean field for the states we are considering. Furthermore it is coupled to a particle reservoir which supplies a chemical potential $\mu S_z$. Finally, it is in interaction with a superconductor which imprints its phase by a term $\nu S_B$ --- the component of $S$ in suitable direction to be determined later. Thus we have a 4-dimensional parameter space $\cP$, two coupling constants -- $\lambda_B$ and $\lambda_M$, and two chemical potentials. For each point in $\cP$ there is a transition temperature $T_c(\lambda_B, \lambda_M, \mu, \nu)\geq 0$ and we want to exhibit a region $\R\in\cP$, where this function is not bounded, that is to say that $T_c$ in this region can become arbitrarily high.

In formulae we start with the Hamiltonian
\beq
H = -\frac{\lambda_B}{2N}S_x^2 -\frac{\lambda_M}{2N}S_z^2 -\mu S_z - \nu S_B,
\eeq
which in the mean-field regime becomes
\beq
H = -\lambda_B S_x \frac{\langle S_x \rangle}{N}-\lambda_M S_z \frac{\langle S_z \rangle}{N} -\mu S_z - \nu S_B.
\eeq

$S_B$ is the combination of $S_x$ and $S_z$ which prevails in the superconducting reservoir and should be a perfect match of the combination we are getting for our system. This sounds somewhat mysterious, as such a cloning seems to require some foresight from the external system. However, there is some redundance in the coefficient of $S_z$ which has contributions both from $\mu$ and $\nu$. We are supposed to be able to control $\mu$ and by giving part of it to $\nu$, we can use it to adjust the direction of $S_B$ so that it coincide with the direction we shall obtain. Thus we write
\beq
H =  (W+\nu) \left[\dfrac{\lambda_B\langle S_x\rangle}{NW}S_x  +
\frac{1}{W}\left(\mu -
\dfrac{\lambda_M\langle S_z\rangle}{N}\right)S_z\right],
\eeq
with
\beqa
W = & \sqrt{ \dfrac{\lambda_B^2\langle S_x\rangle^2}{N^2}
+ \left(\mu-\dfrac{\lambda_M\langle S_z\rangle}{N}\right)^2} \no\\[6pt]
& =: \sqrt{\mu_{eff}^2 +
\dfrac{\lambda_B^2\langle S_x\rangle^2}{N^2}} \geq |\mu_{eff}|.
\eeqa

This Hamiltonian gives rise to the following system of coupled gap-equations:
\beqa
\dfrac{\langle S_x \rangle}{N} &=&
\dfrac{\lambda_B\langle S_x\rangle}{NW}\tanh\dfrac{W+\nu}{2T}\\[8pt]
\dfrac{\langle S_z \rangle}{N} &=&
\dfrac{\mu_{eff}\langle S_z\rangle/N}{W}\tanh\dfrac{W+\nu}{2T}.
\eeqa

The system (2.10--11) has both solutions, corresponding to normal and to superconducting
phases, so with $\langle S_x\rangle = 0$, resp. $\langle S_x\rangle \not= 0$.
In the latter case, from Eq.(2.10) the transition temperature at which the gap closes, $\langle S_x \rangle \rightarrow 0$, is found to be

\beq
T_c = \frac{|\lambda_B|}{2}\,\frac{|\nu+\mu_{{\rm
eff}}|}{|\mu_{{\rm eff}}|}.
\eeq

\noindent As $\mu_{eff}$ can be made arbitrarily small, this means
that for given values of the parameters $\mu$, $\nu$, $\lambda_M$ and $\lambda_B$, in a bounded region $\R\in{\cP}$, the critical temperature can become arbitrarily high, as suggested in
\cite{NPB}.

The second (coupled) gap-equation provides a relation between the model
parameters that determines the relevant parameter range:

$$
\dfrac{\langle S_z \rangle}{N} = \frac{\mu}{\lambda_B+\lambda_M}.
$$

The existence of further order parameters is of severe importance for the
physical content of the models under consideration \cite{Sach}.
Even in the simple model above, the presence of a
second order parameter leads to an enrichment of the structure and to new
effects.

\begin{figure}[ht]
\setlength{\unitlength}{1pt}
\begin{picture}(230,160)
\put(100,-10){\makebox(0,0)[lb] 
{\includegraphics*[90,0][320,150]{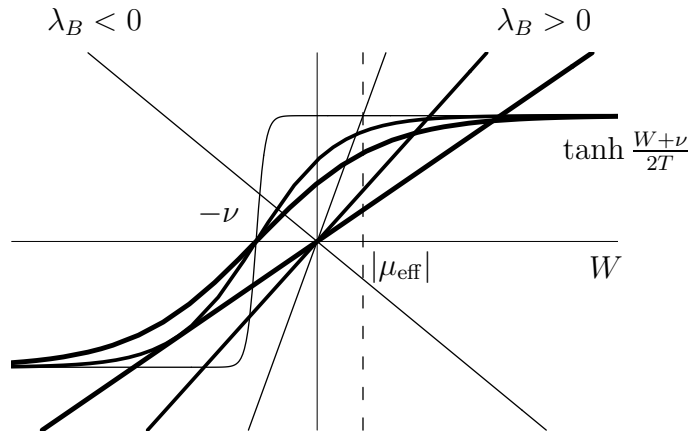}}}
\put(172,75){\makebox(0,0)[l]{$-\nu$}}
\put(237,55){\makebox(0,0)[l]{$|\mu_{\rm eff}|$}}
\put(320,55){\makebox(0,0)[l]{$W$}}
\put(115,148){\makebox(0,0)[l]{$\lambda_B<0$}}
\put(285,148){\makebox(0,0)[l]{$\lambda_B>0$}}
\put(310,99){\makebox(0,0)[l]{$\tanh\frac{W+\nu}{2T}$}}
\end{picture}
\begin{quote}
\caption{Mean-field enhanced BCS: spin-coaxial exchange (plot of both sides
of Eq.(2.15);
the line thickness increases with $T$ and $\lambda_B$).}
\end{quote}
\end{figure}

Let us discus the superconducting solution more in detail.
In the mean-field enhanced model, for $\langle S_x \rangle\not=0$, Eq.(2.10)
reduces to

\beq
\frac{W}{\lambda_B}= \tanh\frac{W+\nu}{2T}.
\eeq
We have chosen the positive eigenvalues of $H$, Eq.(1.5). This is not really a restriction,
since the consideration of the opposite situation will give the conjugate picture.  Thus,
$\tanh (...)$ and $\lambda_B$ must always have the same sign. Also, Eq.(2.9), the lower bound
for the values of $W$ is determined through the effective chemical potential in the
$z$-direction, $\mu_{\rm eff}$. Depending on the coupling of the system to the reservoir (the value
and the sign of $\nu$), we are led to the following situations:

\vspace{0.5cm}
{\bf (A)} \Un{Spin-coaxial exchange, $\nu > 0$} (Figure 4)

\begin{itemize}
\item the BCS-coupling has to be attractive and stronger than the effective
chemical potential, $\lambda_B>|\mu_{\rm eff}|$;
\item the solution (when existing) is uniquely determined;
\item as also seen from Eq.(2.13), the higher-temperature solutions require also stronger
BCS coupling (the thickness of the lines increases with $T$, resp. with $\lambda_B$; the admissible solutions have to be to the right of the dashed line).
\end{itemize}

\begin{figure}[htb]
\setlength{\unitlength}{1pt}
\begin{picture}(450,140)
\put(40,-10){\makebox(0,0)[lb] 
{\includegraphics*[90,0][550,150]{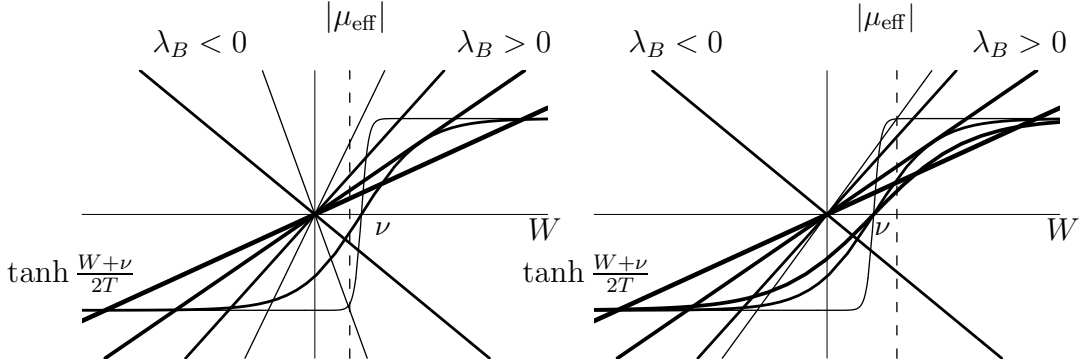}}}
\put(145,125){\makebox(0,0)[l]{$|\mu_{\rm eff}|$}}
\put(346,125){\makebox(0,0)[l]{\small $|\mu_{\rm eff}|$}}
\put(353,45){\makebox(0,0)[l]{\small $\nu$}}
\put(164,45){\makebox(0,0)[l]{\small $\nu$}}
\put(80,115){\makebox(0,0)[l]{$\lambda_B<0$}}
\put(195,115){\makebox(0,0)[l]{$\lambda_B>0$}}
\put(260,115){\makebox(0,0)[l]{$\lambda_B<0$}}
\put(390,115){\makebox(0,0)[l]{$\lambda_B>0$}}
\put(222,45){\makebox(0,0)[l]{$W$}}
\put(418,45){\makebox(0,0)[l]{$W$}}
\put(25,28){\makebox(0,0)[l]{$\tanh\frac{W+\nu}{2T}$}}
\put(220,28){\makebox(0,0)[l]{$\tanh\frac{W+\nu}{2T}$}}
\end{picture}
\begin{quote}
\caption{Mean-field enhanced BCS: spin-anticoaxial exchange.
(a) $|\mu_{\rm eff}|<|\nu|$; (b) $|\mu_{\rm eff}|>|\nu|$ (also here, the line thickness
increases with $T$, resp. with $\lambda_B$; the admissible solutions have to be
to the right of the dashed line)}
\end{quote}
\end{figure}

{\bf (B)} \Un{Spin-anticoaxial exchange, $\nu < 0$}

\bigskip
In this case the relative values of $|\mu_{\rm eff}|$ and $|\nu|$ become of importance.

\medskip
When $|\mu_{\rm eff}|<|\nu|$, (Figure 5a),
\begin{itemize}
\item solutions with repulsive BCS-coupling are possible and uniquely defined;
\item $|\lambda_B|$, when $\lambda_B<0$, has to dominate the effective
chemical potential, $|\mu_{\rm eff}|$;
\item in the positive $\lambda_B$-coupling range, it can happen that
the full system has none, one or two solutions.
\end{itemize}
\smallskip

When $|\mu_{\rm eff}|>|\nu|$, (Figure 5b),
\begin{itemize}
\item only solutions with positive BCS-couplings are possible;
\item depending on the relations between the parameters --- $\mu_{\rm eff}$,
$\lambda_B$ and $\nu$ --- encoded in the second gap-equation, the system can have
none, one or two solutions.
\end{itemize}

As Eq.(2.12) requires small values of $|\mu_{\rm eff}|$ in order to achieve
high transition temperature, this would correspond rather to the situations
depicted on Figures 4 and 5a.

\bigskip

\section{Conclusions}

We considered a BCS-type model with two order parameters, whose solvability
is encoded in two coupled gap equations and which exhibits a high-temperature
superconducting phase. High-$T_c$ superconductivity models that are based
on coupled
gap equations, are known in the literature: such an approach is the one due
to Eliashberg
\cite{E}, see also \cite{Carb} for recent analysis. There, the limitations
on $T_c$ also disappear and are thus interpreted as artifacts of the
Bogoliubov method. However we could not identify the underlying mechanism
with the one described above. Some argumentation for a higher, compared to
BCS, or unlimited transition temperature comes also from the line of
considerations towards an unification of the BCS and BEC pictures
\cite{BEC}. Our model might be relevant here as well as it exhibits an
off-diagonal long-range order, ODLRO \cite{Y, S}. Recall that its existence is the basis
for the Bose--Einstein condensation, however we are dealing here with a fermion
system, so the model provides a framework for analysis of BEC in a Fermi
gas \cite{BECf1, BECf2}.

\bigskip

\section{Acknowledgements}
We thank A. Alekseev, M. B\"uttiker, V. Cheianov, E. Schachinger,
E. Sukhorukov and H.W. Weber for the discussion and H. Narnhofer for useful remarks.
NI thanks Erwin Schr\"odinger Institute for the hospitality. This work
was supported in part by Swiss National Science Foundation.

\bigskip


\end{document}